\documentclass[a4paper,11pt]{article}
\usepackage{pos}
\usepackage{aas_macros}
\usepackage{siunitx}
\usepackage[UKenglish]{babel}
\usepackage{subcaption}

\DeclareSIUnit\year{yr}
\DeclareSIUnit\parsec{pc}
\DeclareSIUnit\gauss{G}

\title{Investigating the CREDIT history of supernova remnants as cosmic-ray sources}

\author*[a]{Anton Stall}
\author[a]{Chun Khai Loo}
\author[a]{Philipp Mertsch}

\affiliation[a]{Institute for Theoretical Particle Physics and Cosmology (TTK), RWTH Aachen University, \\
52056 Aachen, Germany}

\emailAdd{stall@physik.rwth-aachen.de}
\emailAdd{khai.loo@rwth-aachen.de}
\emailAdd{pmertsch@physik.rwth-aachen.de}

\abstract{Supernova remnants (SNRs) have long been suspected to be the primary sources of Galactic cosmic rays. Over the past decades, great strides have been made in the modelling of particle acceleration, magnetic field amplification, and escape from SNRs. Yet while many SNRs have been observed in nonthermal emission in radio, X-rays, and gamma rays, there is no evidence for any individual object contributing to the locally observed flux. Here, we propose a particular spectral signature from individual remnants that is due to the energy-dependent escape from SNRs. For young and nearby sources, we predict fluxes enhanced by tens of percent in narrow rigidity intervals; given the percent-level flux uncertainties of contemporary cosmic-ray data, such features should be readily detectable. We model the spatial and temporal distribution of sources and the resulting distribution of fluxes with a Monte Carlo approach. The decision tree that we have trained on simulated data is able to discriminate with very high significance between the null hypothesis of a smooth distribution of sources and the scenario with a stochastic distribution of individual sources. We suggest that this cosmic-ray energy-dependent injection time (CREDIT) scenario be considered in experimental searches to identify individual SNRs as cosmic-ray sources.}

\ConferenceLogo{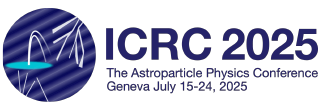}

\FullConference{39th International Cosmic Ray Conference (ICRC2025)\\
 15–24 July 2025\\
Geneva, Switzerland\\}
\begin{document}
\maketitle

%=====================================================================
\section{Introduction}\label{sec:intro}
Cosmic rays (CRs) with energies at least up to the so-called "knee" at $E_{\text{knee}} \simeq \SI{3}{\peta\electronvolt}$ are generally regarded Galactic in origin~\cite{ParticleDataGroup:2024cfk,Gabici:2019jvz,Kachelriess:2019oqu}. 
Among possible source candidates for these Galactic cosmic rays (GCRs), supernova remnants (SNRs)~\cite{2008ARA&A..46...89R} have been ubiquitous in the literature for a long time.
Observationally, many SNRs exhibit power-law spectra across radio~\cite{2019JApA...40...36G}, X-ray~\cite{ChandraSupernovaCatalogue}, and gamma-ray bands~\cite{2016ApJS..224....8A}.
Diffusive shock acceleration of CRs at SNR blast waves can naturally produce such power laws~\cite{2001RPPh...64..429M}.
Furthermore, by converting the kinetic energy of the SNR shock into GCR with an efficiency of $\approx 10\%$, SNRs could account for the locally measured CR energy density~\cite{1964ocr..book.....G}.

However, while SNRs can explain a large part of the CR spectrum, it is not clear if they can accelerate CRs to $E_{\text{knee}}$~\cite{Gabici:2019jvz}.
In~\cite{1983LagageCesarskyA&A}, Lagage and Cesarsky derived the maximum energy of CR particles that can be confined in the source region and thus accelerated.
They found that it is limited to about \SI{10}{\tera\electronvolt} under the assumption of plausible shock speeds and a turbulent magnetic field strength of \SI{1}{\micro\gauss}.
This is much smaller than $E_{\text{knee}}$.
Reaching the "knee" energy would require a strong amplification of the magnetic field compared to the one typical for the interstellar medium.
It has been shown that the so-called non-resonant hybrid (Bell) instability~\cite{2004BellMNRAS,Bell:2013kq}, driven by a current of CRs near SNR shock, can achieve such amplifications early on in the SNR evolution.

In this contribution, we present how the assumption of SNRs as sources for GCRs, together with the consequences of magnetic field amplification in the early stages of the SNR evolution, can lead to imprints on the CR spectrum.
A more detailed discussion of this work can be found in our recent publication~\cite{2025StallLooApJL}.

The idea that CRs of different energies escape their sources in a time-dependent way has been explored for a while in models of CRs in the Galaxy and gamma-ray emissions in the vicinity of CR sources (see, e.g.,~\cite{2009CaprioliBlasiMNRAS2,Schure:2013kya,Bell:2013kq,Cristofari:2020mdf, 2009GabiciAharonianMNRAS,2012ThoudamHorandelMNRAS,2019CelliMorlinoMNRAS}).
It is commonly assumed that escape of CRs from the SNR source region only starts with the beginning of the Sedov-Taylor phase at time $t_{\text{Sed}}$, i.e., once the amount of matter accumulated by the shock is equal to the ejecta mass.
Then, the shock slows down and the magnetic field amplification becomes less effective, leading to a decrease of the maximum energy of CR particles that can be confined in the source region.
This trend can clearly be seen in numerical simulations~\cite{2009CaprioliBlasiMNRAS1}.
Hence, CR particles with higher energies escape the source earlier than low-energy CRs.
The escape time is thus a function of the CR energy, $t_{\text{esc}}(E)$, which connects the escape of CRs of energy $E_{\text{knee}}$ at $t_{\text{Sed}}$ with the escape of all CRs with energies below a threshold energy $E_b$ at the end of the source's lifetime $t_{\text{life}}$.
We refer to this CR escape model as the \emph{Cosmic-Ray Energy-Dependent Injection Time} (CREDIT) scenario.
A more detailed explanation of its implementation will be given in~\ref{subsec: CREDIT} and a discussion of its assumptions can be found in~\cite{2025StallLooApJL}.

Analytical and numerical work on GCR propagation frequently approximates the escaping GCRs with a smooth, steady injection from a source continuum.
However, real SNRs are localized and discrete.
Above \si{\tera\electronvolt} energies, the number of SNRs contributing to the locally measured GCR flux is not necessarily large.
This means that the actual CR flux sourced by discrete SNRs can deviate considerably from the prediction of a smooth model.
However, we only have incomplete knowledge of the coordinates of all CR sources that injected GCRs in the past.
Because of this, Monte Carlo simulations are used to study the statistics of deviations coming from the modelling of discrete GCR sources~\cite{Mertsch2011,Genolini2017,EvoliAmatoBlasi2021_Stochastic,2021PhRvL.127n1101P}.
Especially in the CREDIT scenario, characteristic imprints of individual sources might be observable in the locally measured CR spectrum.

In this work, we illustrate that such spectral structures may indeed arise and can be sizeable — possibly tens of percent.
This makes it worthwhile to explore the detectability of such structures with the current experiments AMS-02~\cite{2021AguilarAliCavasonzaPhR} or DAMPE~\cite{2019AnAsfandiyarovSciA} which report very small statistical uncertainties in their measurements.
We use our Monte Carlo simulations to show that the GCR spectrum produced in a CREDIT scenario can be reliably distinguished from statistical errors on measurements of a spectrum predicted by a smooth source model as well as from a time-independent burst-like injection of sources.
For this, we train a decision tree classifier on a subset of our simulations and evaluate its performance.
These results suggest that analysing high-rigidity CR spectra for discrete spectral bumps might allow us to test the SNR paradigm and GCR escape models directly.
%=====================================================================
\section{Methodology}\label{sec:methodology}
To explore the impact of source discreteness and time dependence on the locally measured CR spectrum, we focus on the proton component of CRs only.
For GCR protons above \SI{10}{\giga\electronvolt}, a simplified diffusion model can be used.
Instead of the proton energy, we use rigidity, defined as $\mathcal{R} = {p c}/{Z e}$ where $p$ denotes the momentum of the particle, $c$ the speed of light, $Z$ the charge number, and $e$ the unit charge.

In this section, we will describe our transport model, the implementation of the CREDIT model, and how we simulate the GCR proton flux, which receives contributions from millions of sources.
The parameters used in the simulations, as well as further details and discussions, can be found in~\cite{2025StallLooApJL}.
\subsection{Cosmic-ray proton transport in the Galaxy}
We assume that CRs propagate in a cylindrical Galactic halo of vertical half-height $H$, with free escape at its boundary at $z = \pm H$.
The isotropic CR density $\psi_\mathcal{R}(\mathcal{R}, t, \mathbf{x})$ ($ = \mathrm{d}n/\mathrm{d}\mathcal{R}$, where $n$ is the number density) evolves according to
\begin{equation}\label{eq:transport_equation}
    \frac{\partial \psi_{\mathcal{R}}\left(\mathcal{R}, t, \mathbf{x}\right)}{\partial t} \;-\; \kappa\left(\mathcal{R}\right) \nabla^2 \psi_{\mathcal{R}} \left(\mathcal{R}, t, \mathbf{x}\right) \;=\; Q\left(\mathcal{R}, t, \mathbf{x}\right) \, ,
\end{equation}
where $\kappa(\mathcal{R})$ is the (isotropic) diffusion coefficient and $Q(\mathcal{R},t,\mathbf{x})$ is the source term describing the CR injection.
We assume that the diffusion coefficient follows a simple power law in rigidity $\kappa(\mathcal{R}) = \kappa_0 \, \beta(\mathcal{R}) \, (\mathcal{R}/\mathcal{R}_{0})^\delta$, with $\beta(\mathcal{R})$ being the velocity in units of the speed of light. 

Other effects like energy losses, reacceleration, and convection can be neglected for protons with rigidities above \SI{10}{\giga\electronvolt}.
We also neglect radial boundary conditions as they influence the flux at the solar position far less than the free escape in the vertical direction for halo heights $H$ of only a few kiloparsecs.
\subsection{CREDIT}\label{subsec: CREDIT}
We now incorporate the time-dependent injection of CRs, following the general notion that more energetic particles escape earlier from SNR shocks.
Each SNR explosion occurs at a random time $t_i$ and position $\mathbf{x}_i$, injecting CRs with an integrated source spectrum $Q_\mathcal{R}(\mathcal{R}) \propto \mathcal{R}^{-\alpha}$.
These coordinates are sampled from an axi-symmetric distribution~\cite{Ferriere2001} in the Galactic disk at $z=0$ with a SNR rate of \SI{0.03}{\per\year}~\cite{Tammann1994_SNRrate}. 

In the CREDIT scenario, the injection time of CRs depends on their rigidity according to $t_{\text{esc},i}(\mathcal{R}) \;=\; t_i \;+\; \Delta t_{\text{esc}}(\mathcal{R})$, where 
\begin{equation}
    \Delta t_{\text{esc}}\left(\mathcal{R}\right) = t_{\text{Sed}} \!\left(\frac{\mathcal{R}}{\mathcal{R}_{\text{knee}}}\right)^a \, , \; a = \frac{\ln\left(t_{\text{life}}/t_{\text{Sed}}\right)}{\ln\left(\mathcal{R}_{\text{b}}/\mathcal{R}_{\text{knee}}\right)} \, ,
\end{equation}
for $\mathcal{R}>\mathcal{R}_{\text{b}}$ and $\Delta t_{\text{esc}}\left(\mathcal{R}\right) = t_{\text{life}}$ otherwise.

Thus, for a single source, the injection term in Eq.~\eqref{eq:transport_equation} is:
\begin{equation}
    Q_i(\mathcal{R}, t, \mathbf{x}) \;=\; 
    Q_{\mathcal{R}}\!\left(\mathcal{R}\right)\;
    \delta\!\bigl(\mathbf{x}-\mathbf{x}_i\bigr)\;
    \delta\!\bigl(t - t_{\text{esc}, i}(\mathcal{R})\bigr).
\end{equation}
Summing over many sources yields the total $Q(\mathcal{R}, t, \mathbf{x})$. 

By contrast, a \textit{burst-like} injection omits the rigidity dependence and sets $t_{\text{esc}, i}(\mathcal{R})$ to a single time for all $\mathcal{R}$. This is often used as a simpler framework in stochastic models~\cite{Mertsch2011,Genolini2017,EvoliAmatoBlasi2021_Stochastic}.

\subsection{Green’s function}
We solve Eq.~\eqref{eq:transport_equation} with the free escape boundary condition for a point source injecting particles of rigidity $\mathcal{R}$ at $(t_i, \mathbf{x}_i)$.
The solution is called the Green's function and can be written as
\begin{equation}
    G\left(t, \mathbf{x}; t_i, \mathbf{x}_i\right) =  \frac{Q_{\mathcal{R}}\left(\mathcal{R}\right)}{\left(2 \pi \sigma^2\right)^{\frac{3}{2}}} e^{-\frac{\left(\mathbf{x}_i-\mathbf{x}\right)^2}{2 \sigma^2}} \vartheta\left(z, \sigma^2, H\right) \, ,
\end{equation}
where $\sigma^2\left(\mathcal{R}, t; t_i \right) = 2 \, \kappa\left(\mathcal{R}\right) \left(t - t_{\text{esc},i}(\mathcal{R})\right)$.
The function $\vartheta$ accounts for free escape boundaries at $z=\pm H$~\cite{Mertsch2011}.
For each supernova, we evaluate the flux at the Sun's position only if the source injection lies within the observer's past light cone~\cite{Genolini2017} to remedy the non-causality of the diffusion transport equation.
Finally, summing contributions from all sources drawn from the source distribution gives a prediction for the local CR proton flux.
\subsection{Computing large ensembles}
Because the transport equation is linear, building up the total flux by summing contributions from randomly drawn SNRs is straightforward, albeit computationally heavy.
We employ GPU acceleration via \textsc{jax}~\cite{jax2018github} to handle tens of millions of discrete sources per realisation efficiently.
The result for each realisation is a spectrum sampled in rigidity bins, which can exhibit significant deviations from a smooth model.
%=====================================================================
\section{Results}\label{sec:results}
\begin{figure*}[!tbh]
\includegraphics[width=\linewidth]{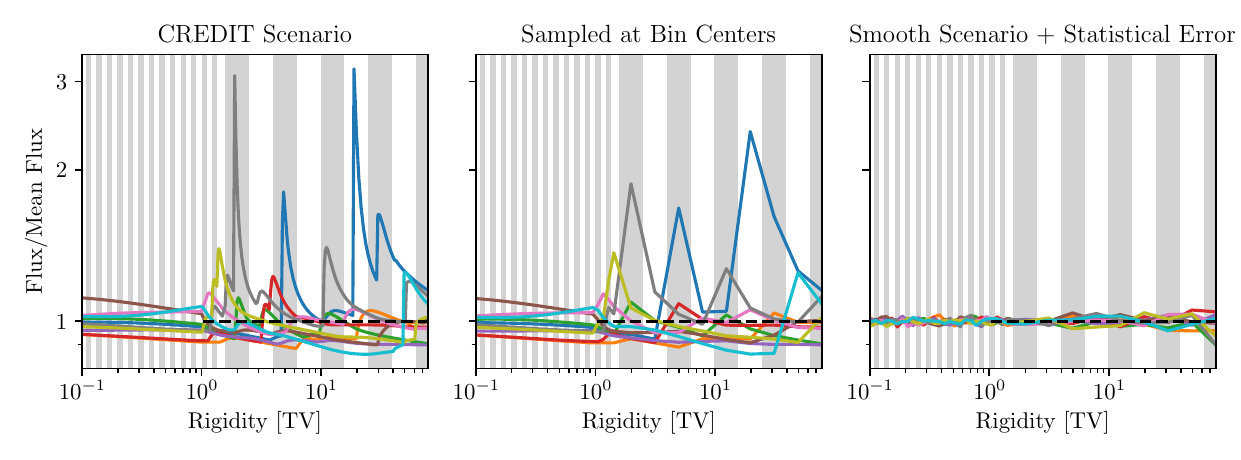}
\caption{GCR proton fluxes (normalized to the ensemble mean) predicted for the CREDIT scenario with $\mathcal{R}_{\text{b}} \in \left[1,25\right]\,\si{\tera \volt}$. The alternating white and gray vertical lines indicate the experiments' rigidity bins.
\textbf{Left:} Ten random realisations, with a fine rigidity resolution. 
\textbf{Middle:} Same spectra sampled at AMS-02 and DAMPE bin centers. 
\textbf{Right:} Smooth model plus uncorrelated 1\% systematic and statistical errors for reference. (This plot is taken from~\cite{2025StallLooApJL}.)}
\label{Fig: panel realisations}
\end{figure*}
In the left panel of Figure~\ref{Fig: panel realisations}, we show example realisations of the CREDIT scenario, displaying notable peaks at rigidities above \SI{1}{\tera\volt}.
These fluctuations can reach tens of percent when a single nearby or young SNR dominates a narrow rigidity range.
The middle panel applies experimental binning representative of AMS-02 and DAMPE.
While somewhat smoothed out, clear enhancements remain visible.
The right panel shows how a smooth and steady injection model might appear once random uncorrelated errors (combining the statistic and a constant systematic error of $1\%$) are added.
It can be seen that the fluctuations in most CREDIT realisations clearly exceed the uncorrelated errors in the right panel.
\subsection{Decision tree classification}
To quantify whether such bumps in the CR spectrum can be reliably recognized, we constructed a large ensemble of simulated proton spectra with $4$~million realisations in each of the following scenarios:
\begin{enumerate}
    \item A \emph{smooth} model with purely systematic plus statistical noise, 
    \item a \emph{burst-like} injection model (stochastic sources, but no rigidity-dependent escape), 
    \item and a CREDIT model with rigidity-dependent injection times. 
\end{enumerate}
We then train a decision tree (DT)~\cite{bishop2006pattern,scikit-learn} on flux values above about $\SI{10}{\giga\volt}$ to see if it can classify each simulation correctly as one of the three scenarios.
\begin{figure}
    \centering
    \begin{subfigure}[b]{0.48\textwidth}
        \centering
        \includegraphics[width=\textwidth]{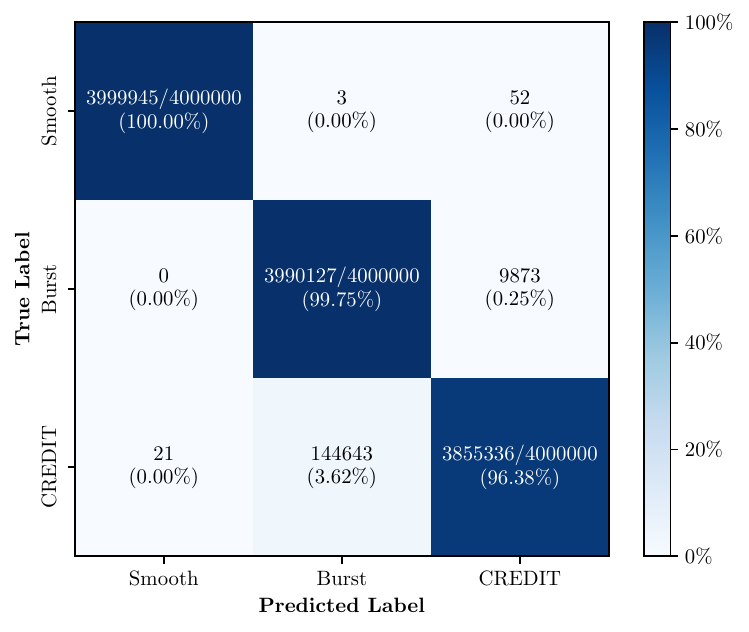}
        \caption{CREDIT realisations are simulated with $20$ distinct, but for each realisation fixed $\mathcal{R}_{\text{b}} \in \left[1,25\right]\,\si{\tera\volt}$.}
        \label{fig2:confusion_matrix_separate}
    \end{subfigure}
    \hfill
    \begin{subfigure}[b]{0.48\textwidth}
        \centering
        \includegraphics[width=\linewidth]{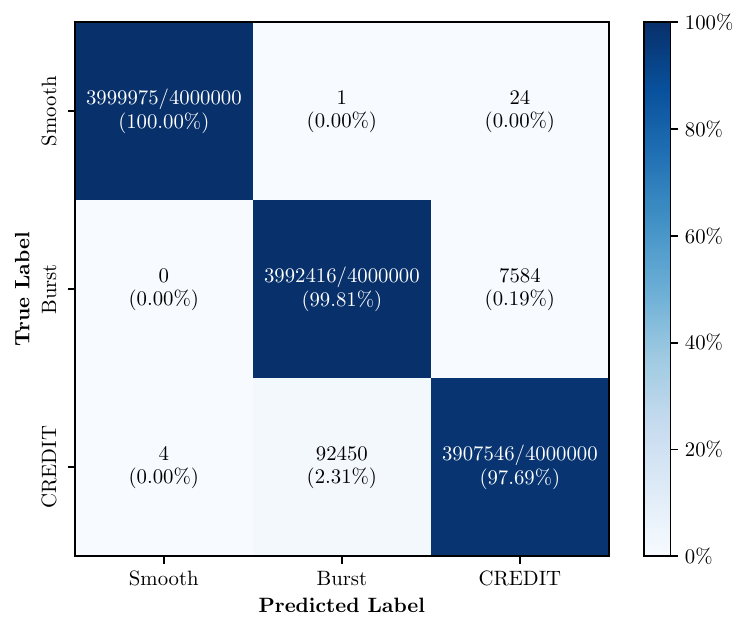}
        \caption{CREDIT realisations are simulated with varying $\mathcal{R}_{\text{b}} \in \left[1,25\right]\,\si{\tera\volt}$ drawn from $p\left(\mathcal{R}_{\text{b}}\right) \propto 1/\mathcal{R}_{\text{b}}$.}
        \label{fig3:confusion_matrix_mixed}
    \end{subfigure}
    \caption{Confusion matrices (normalized to rows for the percentages) for the cross-validation of the DT on the simulation data. (These plots are taken from~\cite{2025StallLooApJL}.)}
    \label{fig:confusion matrices}
\end{figure}
Figures~\ref{fig2:confusion_matrix_separate} and \ref{fig3:confusion_matrix_mixed} show the results of the validation of these DTs in two different choices of how the parameter $\mathcal{R}_{\text{b}}$ is chosen.
They display confusion matrices which show how the label predicted by the DT corresponds to the underlying true label describing the model used to produce the data.
The near-diagonal results demonstrate that the DT successfully separates the three classes reliably.
Hence, narrow bumps in the \si{\tera\volt} range caused by time-dependent CR escape are recognizable above the experimental measurement noise.
In a small number of cases, the two discrete source model realisations are misclassified, but the decision between a smooth and a discrete source model works consistently.
Modifications of various parameters, such as halving the SNR rate or altering the source life time $t_{\text{life}}$ do not strongly affect this conclusion.
We have also checked that spreading the injection over $\sim\SI{1}{\kilo\year}$ around $t_{\text{esc}}(\mathcal{R})$ does not deteriorate the signal for $\mathcal{R}_{\text{b}} \gtrsim \SI{1}{\tera\volt}$.
%=====================================================================
\section{Summary and Conclusion}\label{sec:summary}
We studied the possible imprints of individual SNRs in the locally measured GCR proton flux for the rigidity-dependent escape of CRs from their sources.
This scenario, which we named CREDIT, implements the assumption that high-rigidity protons leave the source region earlier.
We showed how this leads to localised peaks in the \si{\tera\electronvolt} range of the CR spectrum which can be attributed to nearby or recent SNRs.

Using a Monte Carlo approach to model stochastically distributed SNRs, we identified conspicuous features in many realisations.
Crucially, modern experiments’ proton data have an accuracy at or below a few percent; thus, tens-of-percent enhancements in certain rigidity bins are in principle measurable.
A decision tree classification confirms that such features can in principle be distinguished from fluctuations around a smooth model prediction.

If a future dataset displays distinct peaks in the multi-\si{\tera\electronvolt} proton spectrum, it would be an important direct signature of an individual CR accelerator.
Conversely, if the observed flux remains near-smooth within small errors, that outcome poses strong constraints on the SNR paradigm for CR acceleration, especially at high energies.
Ongoing and planned experiments (e.g., AMS-100~\cite{2019NIMPA.94462561S}), extending high-precision coverage up to or beyond \SI{100}{\tera\volt}, are poised to probe these predictions further. 

\acknowledgments
This work has been funded by the Deutsche Forschungsgemeinschaft (DFG, German Research Foundation) — project number 490751943.
The authors also gratefully acknowledge the computing time provided to them at the NHR Center NHR4CES at RWTH Aachen University (project number p0021785).

\bibliographystyle{JHEP}
\bibliography{CREDIT.bib}

\end{document}